\documentclass[preprint,12pt]{elsarticle}
\usepackage{amssymb}
\usepackage[nodots,nocompress]{numcompress}

\begin{document}

\begin{frontmatter}

\title{Detecting multi-spin interaction of an XY spin chain by geometric phase of a coupled qubit}
\author{Xiu-xing Zhang$^{1,2}$}
\author{Ai-ping Zhang$^{1}$}
\author{Fu-li Li$^{1\dag}$}
\ead{flli@mail.xjtu.edu.cn}
\address{$^{1}$MOE Key Laboratory
for Nonequilibrium Synthesis and Modulation of Condensed Matter,
and Department of Applied Physics,
Xi'an Jiaotong University, Xi'an 710049, China\\
$^{2}$Department of Physics, Weinan Normal University, Wei'nan
714000, China}

\begin{abstract}

We investigate geometric phase (GP) of a qubit symmetrically
coupled to a XY spin chain with three-spin interaction in a
transverse magnetic field. An analytical expression for the GP is
found in the weak coupling limit. It is shown that the GP displays
a sharp peak or dip around the quantum phase transition (QPT)
point of the spin chain. Without the three-spin interaction, the
GP has a peak or dip around the critical point $\lambda=1$. If the
three-spin interaction exists, the peak or dip position is
obviously shifted away from the original position. This result
reveals that the GP may be taken as an observable to detect both
the existence and strength of multi-spin interaction in a spin
chain.

\end{abstract}
\end{frontmatter}

\section{Introduction}

Geometric phases (GPs) have been proposed as a typical mechanism
for a quantum system to keep the memory of its evolution in
Hilbert space. Since discovered first by Berry \cite{Berry}, GPs
have become an object of intense research in both the theoretical
and experimental realm \cite{2}. The original concept of GPs has
been generalized to the non-adiabatic case \cite {3}, the
noncyclic case \cite{4} and the degenerate state \cite{7}. It is
also noticed that GPs can be related to a number of important
phenomena in physics \cite{Phenome} such as the Aharonov-Bohm
effect \cite{AB}, the quantum Hall effect \cite{QH} and even
applications in the quantum information processing
\cite{GPC1,GPC2,GPC3,GPC4}.

In recent years, relation between GPs and quantum phase transitions (QPTs)
\cite{QPT} in various closed many-body systems has attracted many interests
\cite{C1,C2,C3,C4}. Carollo \textit{et al. }\cite{C1} showed that GPs are
much sensitive to controlling parameters of spin chains and can be exploited
as a tool to detect critical regions of the systems. Zhu \cite{C2}
investigated GPs in the XY spin chain and showed that the ground-state GP
obeys the scaling behavior in the vicinity of QPTs. And it is also shown
that the connection of GPs with the typical features of QPTs such as the
scaling behavior, critical exponents and so on is not restricted to the XY
spin chain model but universal for quantum many-body systems \cite{C2}.

Since quantum systems are unavoidably to interact with their surroundings,
the time evolution of their states is generally nonunitary. Therefore, it is
desirable to extend the concept of GPs from closed quantum systems to open
ones. Until now, many approaches have been proposed for this purpose \cite%
{5,6,OC1,OC2,OC3}. The first work concerned with this point was given in
Ref. \cite{6}, where GP is investigated purely as a mathematical problem.
Based on the experimental context of quantum interferometry, Sj\"{o}qvist
\textit{et.al } \cite{OC1} introduced a definition of GP for mixed states
undergoing unitary evolution. Tong and coworkers \cite{OC3} developed a
kinematic approach to GP for open quantum systems in nonunitary evolution
led by the environments. According to the definition of GP for open systems,
many works concerned with the correction of environments to GPs have been
done \cite{OC4,OC5,OC6,OC7,OC8}. Yuan \textit{et.al }\cite{OC5} studied GP
of a qubit coupled to an antiferromagnetic spin and found that the GP
changes abruptly to zero when the spin chain undergoes a spin-flop
transition. Lombardo \textit{et.al } \cite{OC6} computed the GP of a
spin-1/2 linearly coupled to a harmonic oscillator reservoir at arbitrary
temperatures and estimated the time scale for experimentally measuring the
GP. Recently, Villar \textit{et.al} \cite{OC7} investigated GP of a spin-1/2
particle in the presence of a composite environment, composed of an external
bath and another spin-1/2 particle. Their results show that the initial
entanglement enhances the sturdiness of GP to decoherence. Cuchietti \textit{%
et.al} \cite{OC8} reported a measurement of GP for a spin-1/2 undergoing
nonunitary evolution induced by coupling of an environment with a NMR
quantum simulator.

In previous studies of GPs with spin environments, only is the
nearest-neighbor spin interaction considered \cite{OC5,YI2}. Besides the
two-spin kind of interactions, however, multi-spin interactions may also
exist in spin chains \cite{multi1,multi2,multi3,multi4}. As a result, phase
transition points may be changed by the multi-spin interaction. In previous
investigations, it has been shown that GPs of a qubit coupled to a spin
chain change greatly around phase transition points of the spin chain and
can signal out the appearance of phase transitions. According to this point,
GPs are expected as a detector of phase transitions of quantum many-body
systems \cite{C1,C2,C3,C4}. With the same reason, we expect that GPs of a
qubit coupled to a spin chain with multi-spin interaction may be used as an
observable to detect the multi-spin interaction. In the present work, we
shall investigate GPs of a qubit symmetrically coupled to an anisotropic XY
spin chain, in which a three-spin interaction is included besides the
nearest-neighbor spin interaction. According to the definition of GP given
in Ref. \cite{OC3}, we obtain an analytical expression for the GP of the
central qubit in the weak coupling limit. Our results show that the
variation of GP at the critical points of quantum phase transitions is much
sensitive to the three-spin interaction, and the GP can be taken as a tool
to detect the existence and strength of multi-spin interaction in spin
chains.

The present paper is organized as follows. In Section 2, the model
is introduced. In Section 3, an analytical expression of the GP is
obtained in the weak coupling limit and the effect of the
multi-spin interaction on the GP is investigated. Finally, a brief
summary is given in Section 4.

\section{The Model}

The model under consideration is composed of a central spin-1/2 (or qubit)
and a $N$-spin-1/2 chain. The central spin is symmetrically and transversely
coupled to the circle spin chain, in which besides the nearest-neighbor spin
interaction a three-spin interaction is also included. Meanwhile, a
transverse magnetic field is homogeneously applied to each spin of the
chain. The associated Hamiltonian reads
\begin{eqnarray}
H &=&\eta \sigma _{0}^{z}-\sum_{l=1}^{N}\left[ \frac{\left( 1+\gamma \right)
}{2}\sigma _{l}^{x}\sigma _{l+1}^{x}+\frac{\left( 1-\gamma \right) }{2}%
\sigma _{l}^{y}\sigma _{l+1}^{y}+\lambda \sigma _{l}^{z}\right]  \nonumber \\
&&-\alpha \sum_{l=1}^{N}\left( \sigma _{l-1}^{x}\sigma _{l}^{z}\sigma
_{l+1}^{x}+\sigma _{l-1}^{y}\sigma _{l}^{z}\sigma _{l+1}^{y}\right) -g\sigma
_{0}^{z}\sum_{l=1}^{N}\sigma _{l}^{z},  \label{Hamiltonian}
\end{eqnarray}%
where $\eta $ is the transition frequency between the ground state $%
(|g\rangle )$ and the excited state $(|e\rangle )$ of the central qubit, $%
\gamma $ and $\alpha $ are the anisotropic parameter of the two-spin
interaction and the strength of the three-spin interaction, respectively, $%
\lambda $ is the coupling constant of the spin chain with the transverse
magnetic field, and $g$ is the coupling strength between the central qubit
and the spin circle. In (\ref{Hamiltonian}), $\sigma _{l}^{m}(m=x,y,z)$ are
the Pauli matrices for spin at the $l$th site of the spin chain and $\sigma
_{0}^{z}(=\left\vert e\right\rangle \left\langle e\right\vert -\left\vert
g\right\rangle \left\langle g\right\vert )$ is the population inversion
operator for the central spin.

The free-motion term of the central qubit in (\ref{Hamiltonian}) can be
removed by the rotating transformation $\exp \left( -i\eta \sigma
_{0}^{z}t\right) $. Then, following the approach proposed in Ref. \cite%
{QPT,multi4}, the rotated Hamiltonian can be fully diagonalized. Since $%
\left[ \sigma _{0}^{z},\sigma _{l}^{m}\right] =0,$ an operator-valued
parameter $\Lambda =\lambda +g\sigma _{0}^{z}$ is a conserved quantity. Thus
$\Lambda $ can be treated as a $c$ number during diagonalizing the
Hamiltonian. Obviously, $\Lambda $ has two eigenvalues $\Lambda _{n}=\lambda
+\left( -1\right) ^{n}g$ with $n=0,1$.

By introducing the Jordan-Wigner transformation \cite{QPT}%
\begin{equation}
\sigma _{l}^{x}=\prod\limits_{i<l}\left( 1-2c_{i}^{\dagger }c_{i}\right)
\left( c_{l}+c_{l}^{\dagger }\right) ,  \label{JW1}
\end{equation}%
\begin{equation}
\sigma _{l}^{y}=-i\prod\limits_{i<l}\left( 1-2c_{i}^{\dagger }c_{i}\right)
\left( c_{l}-c_{l}^{\dagger }\right) ,  \label{JW2}
\end{equation}%
\begin{equation}
\sigma _{l}^{z}=1-2c_{l}^{\dagger }c_{l},  \label{JW3}
\end{equation}%
where $c_{l}^{\dagger }$ and $c_{l}$ are the mapped spinless and fermionic
creation and annihilation operators. Substituting Eqs. (\ref{JW1})-(\ref{JW3}
) into (\ref{Hamiltonian}), one obtains%
\begin{eqnarray}
H &=&-\sum_{l=1}^{N}\left\{ \left[ c_{l}^{\dagger }c_{l+1}+c_{l+1}^{\dagger
}c_{l}+\gamma \left( c_{l}^{\dagger }c_{l+1}^{\dagger }-c_{l}c_{l+1}\right) %
\right] \right.  \nonumber \\
&&\left. +2\alpha \left( c_{l-1}^{\dagger }c_{l+1}+c_{l+1}^{\dagger
}c_{l-1}\right) +\lambda \left( 1-2c_{l}^{\dagger }c_{l}\right) \right\} .
\label{H2}
\end{eqnarray}%
Then by means of the Fourier transformation \cite{QPT}
\begin{equation}
c_{l}=-\frac{1}{\sqrt{N}}\sum_{k=-M}^{M}e^{i2\pi kl/N}d_{k},  \label{eq7}
\end{equation}%
\begin{equation}
c_{l}^{\dagger }=-\frac{1}{\sqrt{N}}\sum_{k=-M}^{M}e^{-i2\pi
kl/N}d_{k}^{\dagger },  \label{equ72}
\end{equation}%
with $M=N/2$ for $N$ even and $M=\left( N-1\right) /2$ for $N$ odd, one gets
\begin{equation}
H=\sum_{k>0}2\left( \Lambda -\cos ka-2\alpha \cos 2ka\right) d_{k}^{\dagger
}d_{k}+i\gamma \sin ka\left( d_{k}^{\dagger }d_{-k}^{\dagger
}-d_{-k}d_{k}\right) ,  \label{H3}
\end{equation}%
where $d_{k}$ and $d_{k}^{\dagger }$\ are the fermionic annihilation and
creation operators in the momentum space, respectively.

In order to diagonalize the Hamiltonian (\ref{H3}), we introduce the
Bogoliubov transformation \cite{QPT}
\begin{equation}
\gamma _{k,\Lambda }=\cos \frac{\theta _{k}^{\Lambda }}{2}d_{k}-i\sin \frac{%
\theta _{k}^{\Lambda }}{2}d_{-k}^{\dagger }.  \label{eq9}
\end{equation}%
By substituting (\ref{eq9}) into (\ref{H3}), the Hamiltonian (\ref%
{Hamiltonian}) can be fully diagonalized and written in the form
\begin{equation}
H=\sum_{k>0}\Omega _{k,\Lambda }\left( \gamma _{k,\Lambda }^{\dagger }\gamma
_{k,\Lambda }-\frac{1}{2}\right) ,  \label{H4}
\end{equation}%
where
\begin{equation}
\Omega _{k,\Lambda }=2\sqrt{\left( \Lambda -\cos ka-2\alpha \cos 2ka\right)
^{2}+\gamma ^{2}\sin ^{2}ka},  \label{eq12}
\end{equation}%
and
\begin{equation}
\theta _{k}^{\Lambda }=\arctan \left( \frac{\gamma \sin ka}{\Lambda -\cos
ka-2\alpha \cos 2ka}\right)  \label{eq13}
\end{equation}%
with $a=2\pi /N$.

It is noted that the fully diagonalized Hamiltonian for the pure spin chain
can be obtained by setting $g=0$ in Eq. (\ref{H4}). Correspondingly, the
energy spectrum $\Omega _{k,\lambda }$, the parameter $\theta _{k}^{\lambda
} $ and the mode operator $\gamma _{k,\lambda }$ for the pure spin chain can
be obtained just by changing $\Lambda $ into $\lambda $ in Eqs. (\ref{eq9})-(%
\ref{eq13}). It can be easily proved that both the mode operators $\gamma
_{k,\lambda }$ for the spin chain and $\gamma _{k,\Lambda }$ for the
qubit-chain coupled system are related by%
\begin{equation}
\gamma _{k,\Lambda }=\gamma _{k,\lambda }\cos \alpha _{k,\Lambda }-i\gamma
_{-k,\lambda }^{\dagger }\sin \alpha _{k,\Lambda }  \label{eq14}
\end{equation}%
with $\alpha _{k,\Lambda }=\left( \theta _{k}^{\Lambda }-\theta
_{k}^{\lambda }\right) /2$.

\section{Geometric Phase of the Central Qubit}

Let us suppose that at $t=0$ the central qubit is decoupled from the spin
environment, and the qubit is in a state
\begin{equation}
\left\vert \phi \left( 0\right) \right\rangle =\cos \frac{\beta }{2}%
\left\vert e\right\rangle +\sin \frac{\beta }{2}\left\vert g\right\rangle .
\label{IS1}
\end{equation}%
The ground state of the spin chain is defined as $\gamma _{k,\lambda
}\left\vert G\right\rangle _{\lambda }=0.$ It can be written in the form $%
\left\vert G\right\rangle _{\lambda }=\prod\limits_{k>0}\left( \cos \frac{
\theta _{k}^{\lambda }}{2}\left\vert 0\right\rangle _{k}\left\vert
0\right\rangle _{-k}+i\sin \frac{\theta _{k}^{\lambda }}{2}\left\vert
1\right\rangle _{k}\left\vert 1\right\rangle _{-k}\right) $, where $%
\left\vert 0\right\rangle _{k}$ and $\left\vert 1\right\rangle _{k}$ are the
vacuum and single excitation states of the $k$th pure environment mode,
respectively. Then according to (\ref{eq14}), $\left\vert G\right\rangle
_{\lambda }$ can be rewritten as
\begin{equation}
\left\vert G\right\rangle _{\lambda }=\prod\limits_{k>0}\left( \cos \alpha
_{k,\Lambda }-i\sin \alpha _{k,\Lambda }\gamma _{k,\Lambda }^{\dagger
}\gamma _{-k,\Lambda }^{\dagger }\right)\left\vert G\right\rangle _{\Lambda
},  \label{GS1}
\end{equation}
where $\left\vert G\right\rangle _{\Lambda }$ is the ground state of the
Hamiltonian (\ref{H4}) and satisfies $\gamma _{k,\Lambda }\left\vert
G\right\rangle _{\Lambda }=0$.

At time $t$, the qubit-environment coupled system unitarily evolutes into
the state $\rho \left( t\right) =U\left( t\right) \left( \left\vert \phi
\left( 0\right) \right\rangle \otimes \left\vert G\right\rangle _{\lambda
\lambda }\left\langle G\right\vert \otimes \left\langle \phi \left( 0\right)
\right\vert \right) U^{\dagger }\left( t\right) $ with the time evolution
operator $U\left( t\right) =\exp \left( -iHt\right) $. By tracing $\rho
\left( t\right) $ over variables of the spin chain, we obtain the reduced
density matrix for the central qubit
\begin{equation}
\rho _{s}\left( t\right) =\left(
\begin{array}{ll}
\cos ^{2}\frac{\beta }{2} & \frac{1}{2}\sin \beta F\left( t\right) \\
\frac{1}{2}\sin \beta F^{\ast }\left( t\right) & \sin ^{2}\frac{\beta }{2}%
\end{array}%
\right) ,  \label{Dem1}
\end{equation}%
where $F\left( t\right) $ is the decoherence factor and is given by
\begin{equation}
F\left( t\right) =_{\lambda }\left\langle G\right\vert U_{\Lambda
_{1}}^{\dagger }\left( t\right) U_{\Lambda _{0}}\left( t\right) \left\vert
G\right\rangle _{\lambda }.  \label{DF}
\end{equation}%
The evolution operators $U_{\Lambda _{0}}\left( t\right) $ and $U_{\Lambda
_{1}}\left( t\right) $ are obtained from $U\left( t\right) $ by replacing $%
\Lambda $ with $\lambda +g$ and $\lambda -g$, respectively. From Eqs. (\ref%
{GS1}) and (\ref{DF}), we can work out the explicit expression for the
modulus of the decoherence factor%
\begin{eqnarray}
\left\vert F\left( t\right) \right\vert &=&\prod_{k>0}\left\{ AB\cos \left(
\Omega _{k,\Lambda _{0}}-\Omega _{k,\Lambda _{1}}\right) t-AB^{2}\sin
^{2}\left( \alpha _{k,\Lambda _{0}}-\alpha _{k,\Lambda _{1}}\right) \right.
\label{DF2} \\
&&\left. +1-\sin ^{2}\left( 2\alpha _{k,\Lambda _{0}}\right) \sin ^{2}\left(
\Omega _{k,\Lambda _{0}}t\right) -\sin ^{2}\left( 2\alpha _{k,\Lambda
_{1}}\right) \sin ^{2}\left( \Omega _{k,\Lambda _{1}}t\right) \right\} ^{1/2}
\nonumber
\end{eqnarray}%
where $A=\sin \left( 2\alpha _{k,\Lambda _{0}}\right) \sin \left( 2\alpha
_{k,\Lambda _{1}}\right) $, $B=2\sin \left( \Omega _{k,\Lambda _{0}}t\right)
\sin \left( \Omega _{k,\Lambda _{1}}t\right) $.

According to the GP definition for an open system, which is proposed in Ref.
\cite{OC3}, the GP acquired by the qubit in a quasi period is given by%
\begin{equation}
\Phi =\arg \left( \sum_{k}\sqrt{\varepsilon _{k}\left( 0\right) \varepsilon
_{k}\left( T\right) }\left\langle \varepsilon _{k}\left( 0\right)
\right\vert \left. \varepsilon _{k}\left( T\right) \right\rangle
e^{-\int_{0}^{T}\left\langle \varepsilon _{k}\left( t\right) \right\vert
\partial /\partial t\left\vert \varepsilon _{k}\left( t\right) \right\rangle
dt}\right) ,  \label{GP2}
\end{equation}%
where $\varepsilon _{k}\left( t\right) $ and $\left\vert \varepsilon
_{k}\left( t\right) \right\rangle $ are the eigenvalues and the
corresponding eigenstates of the reduced density matrix (\ref{Dem1}), and $%
T=2\pi /\eta $ is the time evolution cycle of the qubit when it is isolated
from the environment. If the qubit is coupled to the spin chain, the
evolution of its state is no more periodic. However, if the coupling is
weak, we may consider a quasi cyclic path $P:t\in \left[ 0,T\right] $.

The instantaneous eigenvalues and the corresponding eigenvectors of (\ref%
{Dem1}) are found to be
\begin{equation}
\varepsilon _{\pm }\left( t\right) =\frac{1}{2}\left( 1\pm \sqrt{\cos
^{2}\beta +\sin ^{2}\beta \left\vert F\left( t\right) \right\vert ^{2}}%
\right)  \label{ev2}
\end{equation}%
\begin{equation}
\left\vert \varepsilon _{\pm }\left( t\right) \right\rangle =e^{-i\eta
t}\sin \frac{\beta _{\pm }\left( t\right) }{2}\left\vert e\right\rangle
+\cos \frac{\beta _{\pm }\left( t\right) }{2}\left\vert g\right\rangle ,
\label{ev1}
\end{equation}%
with%
\begin{equation}
\beta _{\pm }\left( t\right) =2\arctan \frac{\cot \beta \pm \sqrt{\cot
^{2}\beta +\left\vert F\left( t\right) \right\vert ^{2}}}{\left\vert F\left(
t\right) \right\vert }.  \label{ev3}
\end{equation}%
Since $\varepsilon _{-}\left( 0\right) =0$, Eq. (\ref{GP2}) shows that only
the eigenvalue $\varepsilon _{+}\left( t\right) $ and corresponding
eigenvector $\left\vert \varepsilon _{+}\left( t\right) \right\rangle $ have
contribution to the GP of the qubit. Upon substituting Eqs. (\ref{GP2})-(\ref%
{ev3}) into Eq. (\ref{GP2}), we obtain the GP of the central qubit
\begin{equation}
\Phi =\eta \int_{0}^{2\pi /\eta }dt\sin ^{2}\frac{\beta _{+}}{2}.
\label{GP3}
\end{equation}

If the qubit has no interaction with the spin chain, i.e. $g=0$, the
well-known result $\Phi =\pi \left( 1+\cos \beta \right) $ can be recovered
from (\ref{GP3}). When the qubit is coupled to the spin chain, the GP will
be modified. In order to get a basic and clear intuition about the effect of
the three-spin interaction on the GP acquired by the qubit in a quasi cycle,
we first make an approximate analysis on the GP.

From Eqs. (\ref{ev3}) and (\ref{GP3}), we see that the correction of the
spin chain to the GP is completely included in the decoherence factor $%
F\left( t\right) $. Therefore, we first analyze the decoherence factor. For
this purpose, following Refs. \cite{Sun}, we introduce a cutoff number $K_{c}
$, which determines the largest energy scale of the spin chain, and define
the partial product
\begin{equation}
\left\vert F\left( t\right) \right\vert
_{c}=\prod_{k>0}^{K_{c}}F_{k}\geqslant \left\vert F\left( t\right)
\right\vert ,  \label{cf}
\end{equation}%
and the corresponding partial sum $S\left( t\right) =\ln \left\vert F\left(
t\right) \right\vert _{c}=\sum_{k>0}^{K_{c}}\left\vert \ln F_{k}\right\vert $%
. For small $k$, we can expand (\ref{eq12}) as a power series of $k$ and
obtain
\begin{equation}
\Omega _{k,\Lambda _{n}}=2\left\vert \Lambda _{n}-1-2\alpha \right\vert
\label{ap1}
\end{equation}%
and%
\begin{equation}
\sin \alpha _{k,\Lambda _{n}}\approx \frac{\left( -1\right) ^{n+1}2\pi
\gamma kg}{N\left\vert \left( \Lambda _{n}-1-2\alpha \right) \left( \lambda
-1-2\alpha \right) \right\vert },  \label{ap2}
\end{equation}%
\begin{equation}
\sin \left( \alpha _{k,\Lambda _{0}}-\alpha _{k,\Lambda _{1}}\right) \approx
\frac{-2\pi \gamma kg}{N\left\vert \left( \Lambda _{0}-1-2\alpha \right)
\left( \Lambda _{1}-1-2\alpha \right) \right\vert }.  \label{ap3}
\end{equation}%
In this way, the partial sum $S\left( t\right) $ can approximately be
written as
\begin{equation}
S\left( t\right) \approx -\frac{1}{2}\frac{\left( 2\pi \gamma g\right) ^{2}}{%
N^{2}\left( \lambda -1-2\alpha \right) ^{2}}\sum_{k>0}^{K_{c}}k^{2}.
\label{ap4}
\end{equation}%
If $K_{c}$ is small, in the thermal dynamical limit $N\rightarrow \infty $,
one may ignore all the terms related to $N^{-l}$ with $l>3$ in (\ref{ap4}).
Consequently, in a short time, we have%
\begin{equation}
\left\vert F\left( t\right) \right\vert _{c}\approx e^{-\tau t^{2}},
\label{ap5}
\end{equation}%
where $\tau =8E\left( K_{c}\right) \gamma ^{2}g^{2}/\left( \lambda
-1-2\alpha \right) ^{2}$ and $E\left( K_{c}\right) =4\pi ^{2}K_{c}\left(
K_{c}+1\right) \times \left( 2K_{c}+1\right) /\left( 6N^{2}\right) $. This
result shows that the decoherence factor would decay in a Guassian-type
behaviour in a short time. Compared with that in the Ising spin environment,
the decay rate $\tau $ is modulated by the three-spin interaction $\alpha $
of the chain \cite{Sun}.

Upon substituting (\ref{ap5}) into (\ref{ev3}), keeping all the terms up to
the second order of the coupling strength $g$ and completing the integral (%
\ref{GP3}), we obtain the approximate expression for the GP
\begin{equation}
\Phi =\pi \left( 1+\cos \beta \right) +\frac{64E\left( K_{c}\right) \pi
^{3}\gamma ^{2}\cos \beta \sin ^{2}\beta }{3\eta ^{2}\left( \lambda
-1-2\alpha \right) ^{2}}g^{2}.  \label{GP4}
\end{equation}%
In the above expression, the first term comes out when the qubit is not
coupled to the spin chain and undergoes a unitary evolution. The second term
is the modification induced by the spin chain. Since the modification is
proportional to the squared coupling strength $g$, and the squared
reciprocal of the transition frequency $\eta$ and the criticality factor $%
\left( \lambda -1-2\alpha \right) ^{-2}$, one may expect that the three-spin
interaction can dramatically change the GP modification around the point $%
\lambda -1-2\alpha =0$ except $\beta =0,\pi /2$ and $\pi $. Meanwhile the GP
modification is positive when $\beta <\pi /2$ and negative when $\beta >\pi
/2$, and changes sign around $\beta =\pi /2$.

In order to get the entire picture about the effect of the three-spin
interaction on the GP, we have numerically investigated the variation of GP
with various parameters of the spin chain according to Eq. (\ref{GP3}).

\begin{figure}[tbph]
\centering \includegraphics[width=9.0cm]{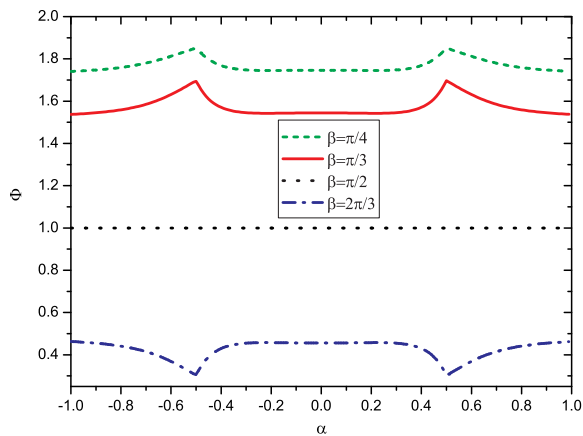} \caption{The GP
versus the three-spin interaction strength $\protect\alpha$ for
the different initial states in the Ising limit ( $\protect\gamma
=1$) and without the external field ($\protect\lambda =0$). The
other parameters are chosen as $N=501, g=0.03$ and $\protect\eta
=2\protect\pi /3$.} \label{fig:Fig1}
\end{figure}

When the transverse magnetic filed is absent, i.e. $\lambda =0$, Eq. (\ref%
{GP4}) shows that the GP modification may dip or peak around the point $%
\alpha =-0.5$, depending on the initial state of the qubit. In Fig. 1, the
GP is plotted as a function of the three-spin interaction strength $\alpha $
for the different initial states in the Ising limit $(\gamma =1)$. The
features of the GP on the parameter $\beta $ shown in Fig. 1 are same as
expected from Eq. (\ref{GP4}). For a fixed value of $\beta $, except the
points $\beta =0,\pi /2$ and $\pi $, the GP displays a peak or dip at the
critical point $\alpha =-0.5$. Moreover, it is noted that the GP also has a
peak or dip at the critical point $\alpha =0.5$ which can not be expected
from Eq. (\ref{GP4}). In Eq. (18), the variable $k$ takes values from $0$ to
$N/2$. Eqs. (11)-(12) depend on the variable $k$ through the trigonometric
functions $\sin (2\pi k/N)$ and $\cos (2\pi k/N)$. Thus, Eqs. (11)-(12) must
have dips or peaks around $k=N/2$ if it has dips or peaks around $k=0$. It
is this reason that leads to the appearance of peaks or dips of the GP at $%
\alpha =0.5$. Fig. 1 also shows that the critical point position does not
change with variation of $\beta$, and is determined only by the three-spin
interaction.

\begin{figure}[tbph]
\centering \includegraphics[width=11.0cm]{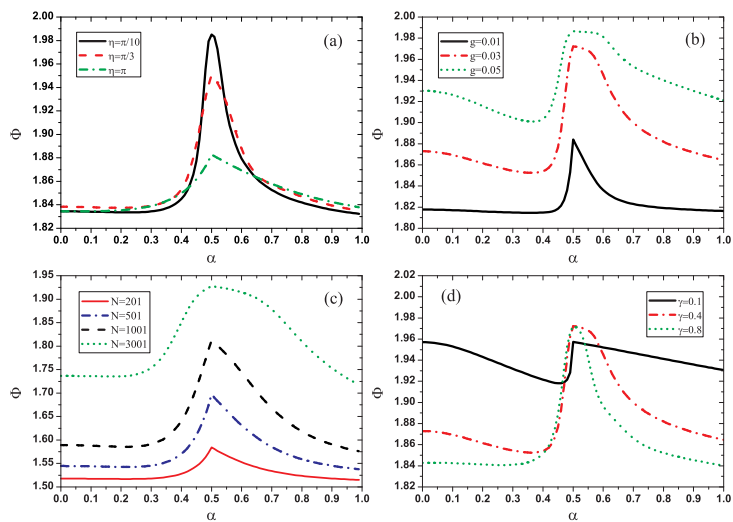} \caption{The
GP versus the three-spin interaction $\protect\alpha$ with (a)
different values of of the transition frequency $\protect\eta$ and $N=501,$ $%
g=0.03$, $\protect\beta =\protect\pi /5$ and $\protect\gamma = 1$; (b)
different values of the qubit coupling strength $g$ and $N=501,\protect\eta =%
\protect\pi /5$, $\protect\beta =\protect\pi /5$ and $\protect\gamma = 0.4$;
(c) different values of the spin chain size $N$, and $g=0.03$, $\protect\eta %
=2\protect\pi /3$, $\protect\beta =\protect\pi /3 $, $\protect\gamma = 1$
and (d) different values of the anisotropic parameters $\protect\gamma$, and
$g=0.03$, $N=501, \protect\eta =\protect\pi /5$, $\protect\beta =\protect\pi%
/5$.}
\label{fig:Fig2}
\end{figure}

In Figs. 2, the GP is plotted as a function of $\alpha$ with various values
of the transition frequency $\eta $, the coupling strength $g$ and the spin
chain size $N$ for the case of $\lambda =0$, respectively. The GP versus $%
\alpha $ is shown in Fig. 2(a). It is observed that the variation of the GP
is sharpened around $\alpha =0.5$ by decreasing the transition frequency $%
\eta $, which is in accordance with the analytical result of Eq. (\ref{GP4}%
). The influence of the coupling strength $g$ on the GP is shown in Fig.
2(b). Obviously, the peaks of the GP become less pronounced with increasing $%
g$. It means that as a tool of detecting QPTs the GP is powerful only in the
weak coupling case. The influence of the spin chain size $N$ on the GP is
given in Fig. 2(c). It is seen that the peak becomes less sharp with
increasing $N$. The GP of the qubit coupled to the XY spin chain($\gamma
\neq 1$) is shown in Fig. 2(d). Obviously, the larger the anisotropy
parameter is, the more sharp the GP peak is.

Combing Fig. 1 with Fig. 2, we see that by varying the parameters the peak
values of the GP is changed but the peak position of the GP is retained.
Thus, the QPT at $\left\vert \alpha \right\vert =0.5$ induced by the
intrinsic three-spin interaction cab be clearly singled out by the GP of the
central qubit.

\begin{figure}[tbph]
\centering \includegraphics[width=9.0cm]{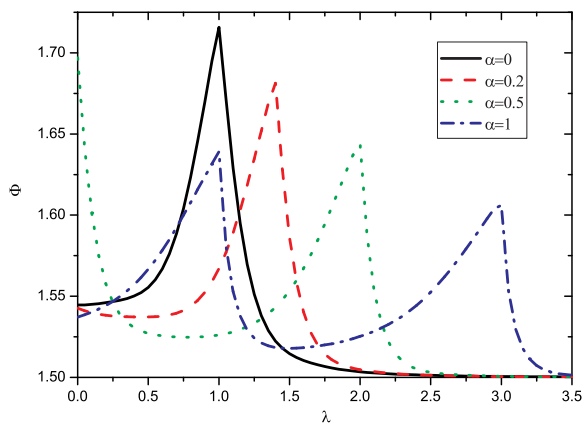} \caption{The GP
as a function of $\protect\lambda $ with different values of the
three- spin interaction strength $\protect\alpha $ in the Ising
limit.
The other parameters are chosen as$\protect\gamma =1$, $N=501$, $g=0.03$, $%
\protect\eta =2\protect\pi /3$ and $\protect\beta =\protect\pi /3$.}
\label{fig:Fig3}
\end{figure}

In Fig. 3, the GP is shown as a function of $\lambda $ with various values
of the three-spin interaction strength $\alpha $ in the Ising limit. When
the three-spin interaction is absent, i.e. $\alpha =0$, the spin chain has a
ground-state phase transition at $\lambda =1$. As shown in Fig. 3, the GP
displays a peak at $\lambda =1$ when $\alpha =0$. When the three-spin
interaction is switched on ($\alpha \neq 0$), as expected from (\ref{GP4}),
the three-spin interaction changes the original critical point and the GP
peak has a $2\alpha $ displacement from the point $\lambda =1$.

\begin{figure}[tbph]
\centering \includegraphics[width=11.0cm]{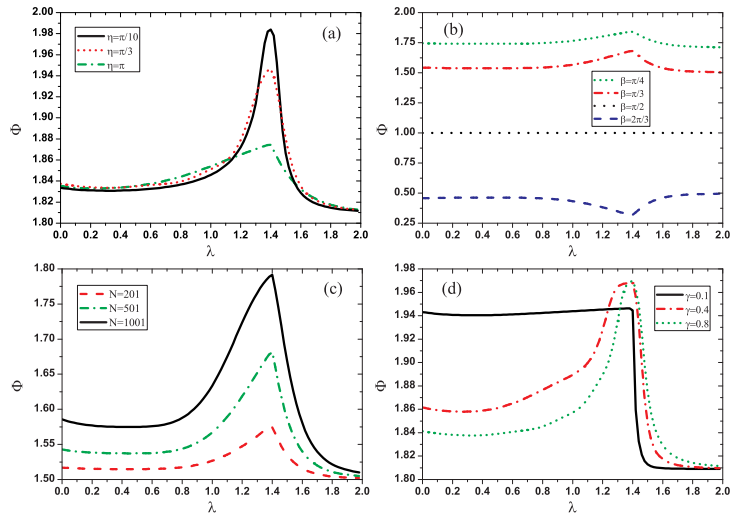} \caption{The
GP versus the coupling constant $\protect\lambda $ for $
\protect\alpha =0.2$ with (a) different values of of the
transition frequency $\protect\eta$ and $N=501,$ $g=0.03$,
$\protect\beta =\protect\pi /5$ and $\protect\gamma = 1$; (b)
different values of $\protect\beta $ and $ N=501, \protect\eta
=2\protect\pi /3$, $g=0.03$ and $\protect\gamma = 1$; (c)
different values of the spin chain size $N$, $g=0.03$,
$\protect\eta =2 \protect\pi /3$, $\protect\beta =\protect\pi /3
$, $\protect\gamma = 1$ and (d) different values of the
anisotropic parameters $\protect\gamma$, and $ g=0.03$, $N=501,
\protect\eta =\protect\pi /5$, $\protect\beta =\protect\pi /5 $.}
\label{fig:Fig4}
\end{figure}

In Figs. 4, the GP is plotted as a function of $\lambda$ with $\alpha =0.2$
for different values of the transition frequency $\eta $, the initial state
coefficient $\beta $, the spin chain size $N$ and the anisotropy parameter $%
\gamma $. In the Ising limit, by setting $N=501$, $g=0.03$, $\gamma =1$ and $%
\beta =\pi /5$, the variation of the GP is shown against $\lambda $ in Fig.
4(a). It is observed that the GP peaks have a $0.4$ displacement from $%
\lambda =1$ and become more sharp around $\lambda =1.4$ with decreasing $%
\eta $, much similar to that in Fig. 2(a). By choosing $N=501,$ $\eta =2\pi
/3$ and $g=0.03$, Fig. 4(b) shows the influence of $\beta $ on the GP. It
can be seen that the GP peaks also move to the point $\lambda =1.4$ but the
peak value decreases with increasing $\beta $. The influence of the spin
chain size $N$ on the GP is shown in Fig. 4(c) with the parameters $%
g=0.03,\eta =2\pi /3$ and $\beta =\pi /3$. It is seen that the GP
peak value increases with increasing $N$. In Fig. 4(d), the GP is
plotted as a function of the anisotropy parameter with $N=501,\eta
=\pi /5$ and $\beta =\pi /5$. It is clear that the GP peak becomes
pronounced but the critical point is not shifted as the anisotropy
parameter increases.

\section{Summary}

We investigate geometric phase (GP) of a qubit symmetrically
coupled to an anisotropic XY spin chain with three-spin
interaction in a magnetic field. An analytical expression for the
GP is obtained in the weak coupling limit. We find that the GP
displays a peak or dip at the quantum phase transition (QPT)
points of the spin chain, depending on the initial state of the
qubit. Without the three-spin interaction, the QPT appears at the
critical point $\lambda =1$ and the GP has a peak or dip around
this critical point. If the three-spin interaction exists, the
peak or dip position of the GP is obviously shifted away from the
original position ($\lambda =1$). Moreover, it is analytically and
numerically shown that the position shift of the GP peak or dip is
determined only by the three-spin interaction. Thus, it comes to
the conclusion that the GP can be used as a tool to detect both
the existence and strength of multi-spin interaction in a spin
chain.

\section*{Acknowledgments}

This work was supported by the National Basic Research Program of China (No.
2010CB923102), Special Prophase Project on the National Basic Research
Program of China (Grant No.2011CB311807) and the National Nature Science
Foundation of China (Grant No. 11074199).

\end{document}